\documentclass[aps,twocolumn,showpacs,superscriptaddress,prc]{revtex4}

\usepackage{epsfig}

\begin{document}
\newcommand{\tb}{ {\bf {t}}}

\newcommand {\Co}{$^{57}$Co}

\newcommand {\Ar}{$^{39}$Ar}

\newcommand {\Na}{$^{22}$Na}

\newcommand {\fP}{$f_{p}$}

\newcommand {\mus}{$\mu$s}

\title{Scintillation time dependence and pulse shape discrimination in liquid argon}
\author{W.~H.~Lippincott}
\affiliation{Department of Physics, Yale University, New Haven, CT 06511}
\author{K.~J.~Coakley}
\affiliation{National Institute of Standards and Technology, Boulder, CO
  80305}
\author{D.~Gastler}
\affiliation{Department of Physics, Boston University, Boston, MA 02215}
\author{A.~Hime}
\affiliation{Los Alamos National Laboratory, Los Alamos, NM 87545}
\author{E.~Kearns}
\affiliation{Department of Physics, Boston University, Boston, MA 02215}
\author{D.~N.~McKinsey}
\email{daniel.mckinsey@yale.edu}
\affiliation{Department of Physics, Yale University, New Haven, CT 06511}
\author{J.~A.~Nikkel}
\affiliation{Department of Physics, Yale University, New Haven, CT 06511}

\author{L.~C.~Stonehill}
\affiliation{Los Alamos National Laboratory, Los Alamos, NM 87545}

\date{\today}

\begin{abstract}

Using a single-phase liquid argon detector with a signal yield of 4.85
photoelectrons per keV of electronic-equivalent recoil energy (keVee), we measure the scintillation time dependence of both electronic and
nuclear recoils in liquid argon down to 5 keVee. We develop two
methods of pulse shape discrimination to distinguish between electronic and
nuclear recoils. Using one of these methods, we measure a background and
statistics-limited level of electronic recoil contamination to be $7.6\times10^{-7}$
between 52 and 110 keV of nuclear recoil energy (keVr) for a nuclear recoil acceptance of $50\%$ with no
nuclear recoil-like events above 62 keVr. Finally, we develop a maximum likelihood
method of pulse shape discrimination 
based on the measured scintillation time dependence.

\end{abstract}


\maketitle

  \section{Introduction}
  \label{sec:intro}
 
Recent years have seen an increase in the number of experiments using noble
liquids as materials for detecting Weakly Interacting Massive
Particles (WIMPs), a well motivated dark matter
candidate~\cite{Jungman:1996}. The current best limit for the spin-independent WIMP-nucleon
cross section for a 60-GeV WIMP mass is $\rm 4.6\times 10^{-44}\,cm^{2}$, set
by the Cryogenic Dark Matter Search (CDMS) experiment~\cite{Ahmed:2008}.  Because many
noble liquids have high scintillation yields, are easily purified of
radioactive impurities, and are likely scalable to large masses with relative
ease, they hold great promise for this application. The current best limit
from a noble liquid detector is $8.8\times 10^{-44}$ cm$^2$ at 100 GeV, set by
XENON10~\cite{Angle:2007}.  Noble liquid
detectors with larger target masses will likely improve on these limits.

 The key to the noble liquid dark matter detectors is discriminating between nuclear recoil events that constitute a WIMP signal
and electronic recoil events that form the primary backgrounds. The
XENON and ZEPLIN experiments are designed to collect both scintillation light
and ionization from liquid xenon~\cite{Aprile:2005a,Cline:2003}. These are
dual-phase detectors that use both scintillation light and
ionization charge collection to discriminate between event classes, as nuclear recoils and
electronic recoils produce different ratios of charge to light. Liquid
argon is an attractive
alternative to liquid xenon due to the lower cost of natural argon and its simpler
purification requirements. The
WIMP Argon Programme (WARP) and Argon Dark Matter (ArDM) experiments employ dual-phase detectors that use liquid
argon as the target~\cite{Rubbia:2005,Brunetti:2005}. 

Alternatively, a single-phase detector collecting solely scintillation light might distinguish electronic and
nuclear recoils using pulse shape discrimination (PSD). In 1977, Kubota \textit{et
al.} showed that the time dependence of scintillation light in liquid xenon
and liquid argon is significantly different for heavy ionizers such as $\alpha$
particles and fission fragments when compared to light ionizers such as
$\beta$ decay and
Compton-scattered electrons~\cite{Kubota:1978}. This is because scintillation in
liquid noble gases is produced by the decay of excimers that can exist in either singlet or
triplet molecular states, which have very different lifetimes (Table
\ref{table:nobles}). The slow scintillation light emitted by triplet molecules
can be suppressed in
intensity by destructive triplet interactions, primarily Penning ionization
and electron-triplet spin exchange; it is believed that these reactions are stronger for high
excitation densities such as those produced by nuclear recoils, causing the
observed time dependences. Therefore, the relative amplitudes of the fast and slow components can be used to
determine which type of excitation occurred for a given event.

\begin{table}[!ht]
\centering
\begin{tabular}{ccc} \hline\hline
 & Singlet Lifetime (ns) & Triplet Lifetime (ns) \\ \hline
\,Ne\, & $< 18.2 \pm 0.2$ & $14 900 \pm 300$ \\ 
\,Ar\, & $7.0 \pm 1.0$ & $1600 \pm 100$ \\ 
\,Xe\,  & $4.3 \pm 0.6$  & $22.0 \pm 2.0$ \\ \hline\hline

\end{tabular}
\caption{Lifetimes of the singlet and triplet states for neon, argon, and
  xenon excimers~\cite{Hitachi:1983,Nikkel:2007}.}
\label{table:nobles}
\end{table}

Pulse shape discrimination based on the timing of scintillation light has been
studied for use with several noble liquids. In liquid helium, PSD has been
studied in order to separate electronic recoil events from $^3$He({\it n,p})$^3$H
events in the search for the permanent electric dipole
moment of the neutron~\cite{McKinsey:2003}. PSD 
has also been used to suppress $\gamma$-ray backgrounds in liquid xenon~\cite{Akimov:2002,Davies:1994}. McKinsey and
Coakley~\cite{McKinsey:2005b} pointed out that the much longer triplet lifetime in liquid neon
should allow superior PSD, which has recently been
verified experimentally~\cite{Nikkel:2007}. Following this observation, Boulay and Hime recognized that the
similar properties of liquid argon could in principle achieve PSD with part
per billion levels of electronic recoil contamination
(ERC)~\cite{Boulay:2006}. ERC is defined to be the
probability of incorrectly classifying an electronic recoil event as a nuclear
recoil event given a particular level of nuclear recoil acceptance. An ERC of
$10^{-8}$ or better is required to perform a competitive WIMP 
search using liquid argon due to the presence of the radioactive isotope
$^{39}$Ar, which produces about 1 Bq per kg of atmospheric argon~\cite{Loosli:1969,Formaggio:2004,Benetti:2007b}.
The WARP collaboration has used scintillation timing in combination with an
ionization signal to reduce electronic recoil
backgrounds in liquid argon~\cite{Benetti:2007}. The Dark Matter Experiment
using Argon Pulse Shape Discrimination (DEAP) has demonstrated a 
background limited ERC of $5\times10^{-6}$ using the DEAP-0 single phase detector
for nuclear recoil energies above 1~MeV~\cite{Boulay:2006b}.
At a given energy, ERC improves exponentially with scintillation light
collection efficiency. For this reason, efficient scintillation light
detection is the primary requirement for performing a sensitive WIMP search
with negligible background at a suitably low energy threshold. 

In this paper we describe measurements of scintillation in liquid argon due to
low-energy nuclear and electronic recoils in the energy range relevant to a
WIMP dark matter search.  We measure the scintillation time
dependence of liquid argon for both event classes. We develop two basic PSD
methods and measure the level of discrimination in our apparatus. Finally, we
use the measured time dependence to develop a maximum likelihood method of PSD.

  \section{Experimental Details}
  \subsection{Detector Design}
  \label{sec:det_design}

\begin{figure}[htbp]
  \centerline{
    \hbox{\psfig{figure=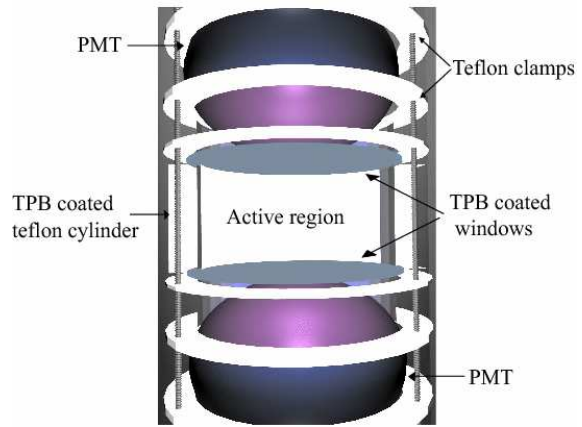,width=7.5cm,%
                clip=}}}
          \caption[Scintillation cell]
                  {(Color online) Schematic representation of the scintillation cell.}
          \label{fig:cell}
\end{figure}

The apparatus consists of a 3.14-liter active volume of liquid argon viewed by
two 200-mm-diameter photomultiplier tubes (PMTs)~\cite{Hamamatsu}, all contained within a stainless steel vessel and
vacuum Dewar. Figure~\ref{fig:cell} shows a schematic of the central volume and
PMTs. The active region is defined by a Teflon cylinder 200~mm in diameter and
100~mm in height with two 3-mm-thick fused-silica windows enclosing the top
and bottom. The PMTs are held in place by Teflon rings above
and below the central volume and view the active region through the
windows. They are powered by positive high voltage with a typical gain of
approximately $4\times 10^{7}$.

 Because liquid argon scintillates in the ultraviolet ($\approx~128
$~nm)~\cite{Cheshnovsky:1972}, all inner surfaces of the Teflon and windows are
coated with a thin film of tetraphenyl butadiene (TPB)~\cite{McKinsey:1997}
by use of a Tectra Mini-Coater evaporation
system~\cite{Tectra}\footnote{Commercial equipment, instruments, or materials
  are identified in this paper to adequately specify the experimental
  procedure. Such identification implies no recommendation or endorsement by
  NIST, nor does it imply that the materials or equipment identified are
  necessarily the best available for the purpose.}. The TPB shifts the wavelength of the ultraviolet light to approximately 440~nm
so that it may pass through the windows and be detected by the PMTs. Both
windows are coated with $(0.20 \pm 0.01)$ mg/cm$^2$ of TPB, while the Teflon
cylinder is coated with $(0.30 \pm 0.01)$ mg/cm$^2$. The Teflon cylinder, windows, and PMTs are all immersed
directly in liquid argon and contained within a 25-cm-diameter by 91-cm-tall
stainless steel vessel. 

The stainless steel vessel is in turn housed inside a vacuum Dewar, and argon
gas is introduced into the system though a tube on the top of the Dewar. The
argon is liquefied in a copper cell mounted to the end of a pulse-tube
refrigerator~\cite{Cryomech} inside the Dewar before flowing
through a tube to the stainless steel vessel. All components
that come into contact with the gas or liquid are baked to at least 60$^\circ$C, and
the ultra-high-purity argon gas (99.999\%) is passed through a heated
gas-purification getter~\cite{Omni} before
entering 
the vessel. In addition, the argon is continually circulated through the
getter and reliquefied at a rate of at least $2.0$~standard liters per minute (slpm) to ensure that high
purity is maintained. The stability of the system is discussed further in
Sec.~\ref{sec:det_calibration}.

\begin{figure}[htbp]
  \centerline{
    \hbox{\psfig{figure=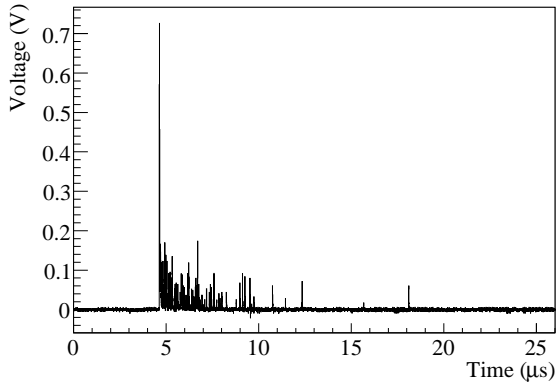,width=8cm,%
                clip=}}}
          \caption[Sample scintillation trace]
                  {Example of an electronic recoil event from a single PMT,
                    digitized by the 8-bit WFD, sampling at 500 MHz. Four
                    microseconds of
                    presamples are recorded to
                  measure the baseline.}
          \label{fig:trace_ex}
\end{figure}

 The data acquisition system is custom-built around VME-bus 
waveform digitizers (WFDs); a sample WFD trace from a scintillation event in argon can be seen in
Fig.~\ref{fig:trace_ex}. The PMT signals from the detector are divided
three ways by a linear fan out with two copies of the signal sent to the
WFDs and one sent to a triggering system. Each WFD has four channels 
that record eight-bit samples at 500~MHz. These samples are stored in a separate 
programmable-length memory buffer for each channel. For all data presented
here, the record length is set to 26~\mus. The two copies of the PMT waveforms are recorded 
separately at unity gain and at an attenuation of ten to increase the effective dynamic 
range of the eight-bit digitization. The buffer is continually filled but saved to
disk only
when the triggering system registers a fraction of a photoelectron in
both PMTs within a 100-ns coincidence
window.  Once a trigger has been registered, the DAQ records the event for
 22~\mus, leaving an additional 4~\mus\, of baseline
presamples in the data~(see Fig.~\ref{fig:trace_ex}).  The data are read to
a computer via fiber-optic cable, and after the computer has recorded all 26~\mus\, of
data, it resets the system for the next event. The data collected by
the DAQ software are saved in a ROOT-based file structure~\cite{Brun:1997}.

  \subsection{Data Collection}
  \label{sec:data_collection}

All data collected are processed in software. First, for each PMT, the two gain scales are
combined into a single waveform. Three~\mus\, of baseline presamples are
averaged to obtain a baseline and baseline root-mean-square. The baseline is then subtracted from
the trace. Ideally, we would count single photoelectrons in a pulse, but since
we detect many photoelectrons that produce signals overlapping in time, we
integrate the trace in order to
determine the total number of photoelectrons. To mitigate the integration noise, we restrict the
range of the integral to 50-ns regions in which the trace voltage crosses a threshold of 4
times the baseline root-mean-square (a threshold of approximately 2/5
the height
of a photoelectron). This method is a hybrid of single
photoelectron counting and pulse integration. 

We then apply three cuts to all data. One cut removes events in which
either PMT saturates due to excessive light exposure by rejecting events
above an experimentally determined threshold of approximately 2000 times the single
photoelectron pulse area. A second cut removes events for
which the trigger time (defined as the time at which the voltage rises above 20\% of its maximum
value) differs by greater than 20 ns between the two PMTs. The
third cut is designed to eliminate events that produce light in the windows or the glass of
the PMTs. An asymmetry parameter $A$ is defined as
\begin{equation}
A = \frac{S_T - S_B}{S_T + S_B},
\end{equation}
where $S_T$ and $S_B$ are the signal areas in the top and bottom PMTs. For most
data, we require $-0.275 < A < 0.375$, due to a slightly larger gain in the
top PMT. This cut is relaxed to $|A| < 0.4$ to improve the statistics of
the nuclear recoil data described below. 

We use a 10-$\mu$Ci \Na\, source to produce
electronic recoils. In 90\% of \Na\, decays, a positron is emitted that
immediately annihilates in the surrounding materials to produce 511-keV $\gamma$
rays with equal and opposite momenta, or ``back-to-back''. We use the second
$\gamma$ ray to tag electronic recoil events in the liquid by triggering on a
coincidence within a 100-ns window between the PMTs in the argon and a NaI
crystal scintillator placed back-to-back with our apparatus. This event tagging
reduces backgrounds in our data from other radioactive decays and cosmic rays. To further decrease neutron backgrounds, we place one layer of water-filled containers above and
around the sides of the dewar. These containers are cubes of side 30 cm in
length and hold 20 liters of water. 

In addition to the three universal cuts, we apply two additional cuts to ensure data
purity. In software we narrow the coincidence window between the liquid argon PMTs and the
NaI crystal to 30 ns. We also make an
energy cut to the NaI energy spectrum to select 511-keV events in the NaI crystal. In the liquid
argon, the
majority of the 511-keV $\gamma$s Compton scatter, producing
events with a continuum of deposited energies.

\begin{figure}[htbp]
  \centerline{
    \hbox{\psfig{figure=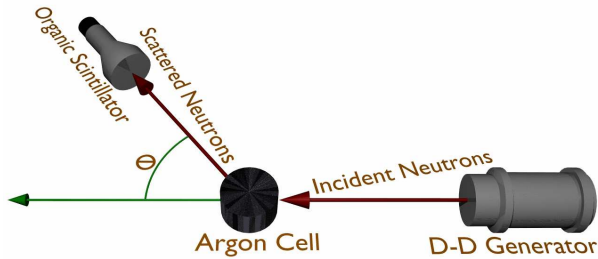,width=8cm,angle=0,
                clip=}}}
          \caption{(Color online) Schematic of the neutron scattering setup.  
                  }
          \label{fig:nuc_scatt}
\end{figure}

To investigate the detector response to nuclear recoils, we use a portable
deuterium-deuterium neutron generator~\cite{Thermo} as a source of 2.8-MeV neutrons and a PMT viewing BC501A organic scintillator as a secondary
detector. Both the generator and the organic scintillator are placed approximately
1.63~m from the center of the active volume. We require a detection of a scintillation event in the liquid argon,
followed within 200 ns by an event in the organic scintillator. The experimental setup can be seen schematically in
Fig.~\ref{fig:nuc_scatt}. By changing the scattering angle $\theta$, we can choose the energy of the nuclear recoils
observed in the liquid, $E_{rec}$, using simple kinematics:
\begin{eqnarray}
 \nonumber E_{rec} =&
\frac{2E_{in}}{(1+M)^2}[1+M-\cos^2(\theta) \\
&-\cos(\theta)\sqrt{M^2+\cos^2(\theta)-1}],
\end{eqnarray}
where $E_{in}$ is the incident neutron energy and $M$ is the atomic mass of the
target.

Since we know the energy of the neutron, we can
calculate the time-of-flight of a neutron that scatters in the liquid
argon and the organic scintillator. We then apply a time-of-flight cut between the liquid
argon cell and the organic scintillator to distinguish between $\gamma$ rays and
neutrons, as the neutron generator produces both. In general, the
time-of-flight cut requires an event to occur in the organic scintillator
60--90 ns after the event in the liquid argon. In addition, we make a PSD cut
in the organic scintillator data to further eliminate electronic recoils.

We also collect background data by looking at events in the
liquid argon with no external source present. This background data provides an
estimate of the accidental background rate that may be
contaminating the $\gamma$-ray data sets, and it will 
be discussed further in Sec.~\ref{sec:PromptFraction}

  \subsection{Detector Calibration}
  \label{sec:det_calibration}

We use a 10-$\mu$Ci sealed $^{57}$Co source for daily
calibrations. This source produces 122-, 137-, and 14.4-keV $\gamma$ rays, with branching ratios of 86\%, 11\% and 9\%,
respectively. Any scintillation event in liquid argon produces a
significant triplet component; since this component is spread out over many
\mus, it appears in the signal as many single photoelectrons well
separated in time (for example, there are a number of single photoelectrons
that appear after 7~\mus\, in Fig.~\ref{fig:trace_ex}). Therefore, we measure the gain of the PMTs using the
$^{57}$Co source by selecting single photoelectrons from the tail end of
each pulse. The PMT traces are divided into 75-ns regions centered on times at
 which the trace
crossed an experimentally determined threshold of roughly 1/3 of a
photoelectron. These regions are then integrated to obtain the single
photoelectron pulse
area. The typical gain for the PMTs is approximately $4\times10^{7}$.

\begin{figure}[hbtp]
  \centerline{
    \hbox{\psfig{figure=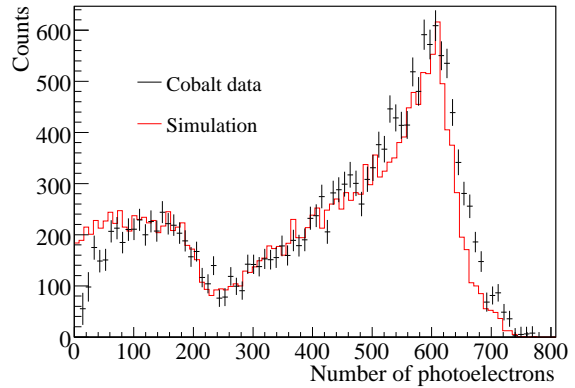,width=8cm,%
                clip=}}}
          \caption[Plot of $^{57}$Co data]
                  {(Color online) Example spectrum of $^{57}$Co data, along with a simulation
                  done using the Reactor Analysis Tool~\cite{RAT}.}
          \label{fig:CoSim}
\end{figure}

Figure~\ref{fig:CoSim} shows an example $^{57}$Co spectrum along with a
simulation done with the Reactor Analysis Tool, a toolkit of Geant4
developed by the Braidwood collaboration~\cite{Agostinelli:2003,RAT}. From the simulation, we find that the
position of the primary peak is dominated by the 122-keV $\gamma$
photoabsorption process. The simulation parameters describing
absorption and reflection of the materials in the detector are tuned to
match the observed signal yield, and the spectral shapes line up nicely. 

By comparing the integrated signal corresponding to the 122-keV peak to that
of a single photoelectron, we measure the signal yield of the detector to be 4.85
photoelectrons per keV electron equivalent (keVee), where keVee refers to the
amount of energy deposited by an electronic recoil. The response of the
detector was
stable to within 5\% during the four months of data acquisition.

We use the 122-keV 
$^{57}$Co peak to provide a daily energy calibration. To check the quality of
that 
calibration, we use the 511-keV $\gamma$ rays produced by the $^{22}$Na source as
a second point of reference. When calibrated using the $^{57}$Co source and the
assumption that the signal scales linearly with deposited energy, the 511-keV
absorption line appears in the $^{22}$Na spectrum as expected to within
1\%. In addition, simulations of 511-keV $\gamma$ rays are consistent with the
data. 

Impurities in the detector can build up over time via outgassing. These impurities can quench argon excimers or absorb emitted UV
photons, which would lead to a decrease in light yield. Additionally, one
would observe a decrease in the triplet molecule lifetime. Work by Himi~{\it et al.} suggests that an impurity level of 0.5 atoms of nitrogen per $10^6$
atoms of argon in the liquid
could decrease the observed triplet lifetime by as much as 0.1
\mus~\cite{Himi:1982}. Further experiments quantifying the reduction of the triplet
lifetime due to nitrogen and oxygen
impurites have recently been performed by the WARP
collaboration~\cite{Acciarri:2008a, Acciarri:2008b}. To avoid signal degradation, we continually circulate the argon through a getter
before reliquefying back into the detector. We use daily measurements of both the light yield and the triplet
lifetime to monitor the purity level. 

\begin{figure}[!htbp]
  \centerline{
    \hbox{\psfig{figure=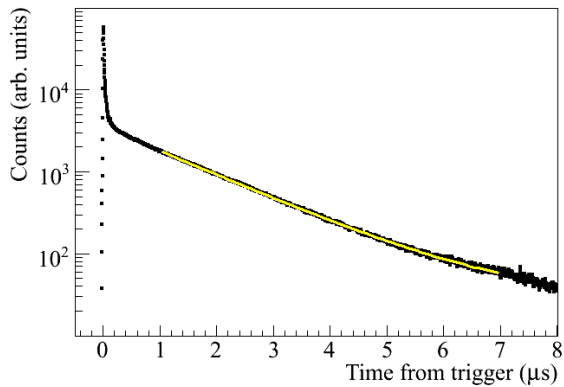,width=8cm,%
                clip=}}}
          \caption{(Color online) An example fit of a single exponential to an average trace to
            measure the long time constant in liquid argon.}
          \label{fig:LTC_fit}
\end{figure}

We measure the light yield in the manner described earlier in this section, and we use the same \Co\, data to measure the triplet
lifetime. First, we select events in the 122-keV peak to make sure we use a similar
data set for each individual measurement. The top and bottom PMT traces are normalized by the size of the
single photoelectron and summed together. We align each pulse based on its
estimated trigger time, defined as
the time at which the trace first crosses
20\% of its maximum value. At this trigger time, the relative time for each
pulse is $t=0$. Between 5000
and 10000 traces are averaged and
the following model is fit to the average trace between 1 and 7
\mus\, from the trigger:
\begin{equation}
<V(t)> = A \exp(-t/\tau_l) + B,
\end{equation}
where $<V(t)>$ is the expected trace, $A$ is a normalization factor, $\tau_l$
is the triplet lifetime and $B$ is an additional baseline term that helps
stabilize the fit over a range of fit windows. An example fit is shown in Fig.~\ref{fig:LTC_fit}.

We find two significant sources of systematic error in the fit parameters stemming from the voltage
applied to the PMTs and the choice
of fit window. Although we do not fully
understand the PMT voltage effect, we estimate the systematic uncertainty to
be 33 ns by
changing the PMT voltages by $\pm 75$ V (equivalent to dividing or multiplying the gain by
2). We do not
fit the data before 1 \mus\, or after 9 \mus, as the fit becomes less reliable
due to contamination from the fast component and baseline noise,
respectively. We estimate the systematic uncertainty associated with the fit
window to be about 35 ns by both varying the
end time by $\pm 2$ \mus\, and by fitting the data within 5 \mus\,
windows ending at 6, 7 and 8 $\mu$s from the trigger.  We combine the two sources of error into a single systematic uncertainty estimate of
50 ns.

During the four months of operation, we measure the long time
constant and signal yield to be $(1463~\pm~5_{\mathrm{stat}}
\pm~50_{\mathrm{sys}})$ ns and $(4.85~\pm~0.08)$ 
photoelectrons/keVee. The uncertainty on the signal yield is statistical, and varying the PMT
voltage by $\pm 75$ V has no apparent effect. For a fixed fit window and PMT
voltage, the long time constant is stable to within $1\%$, while the signal yield is stable to within $5\%$.

  \section{Experimental Results}

  \subsection{Detection Time PDF Model}
  \label{sec:TimePDF}

We measure the time dependence of scintillation light produced by
electronic and nuclear recoil scattering events
in our detector. For an event with energy deposited at time $t_0$, we model the temporal probability density functions (PDFs) for the emission
times of scintillation photons as the weighted sum or mixture of two exponential PDFs:
\begin{equation}
f(t-t_0) ~= q~ g(t-t_0,\tau_l)  + (1-q) ~ g(t-t_0,\tau_s),    
\label{Eq:PDF}
\end{equation}
where
\begin{equation}
g(t,\tau)
=
\frac{1}{\tau}\exp\left(\frac{-t}{\tau}\right),
\end{equation}
 $\tau_l$ and $\tau_s$ are the long and short time constants for both nuclear and electronic
recoil event classes at any given energy, and 
the probability parameter $q$ takes different values for the two classes of events.

At each PMT, detected scintillation photons yield photoelectrons 
that produce observed voltage traces. For the small detector in this experiment,
scintillation transit times are negligible. We assume that the duration of energy deposition
and excimer formation is instantaneous compared to the time scales
relevant to scintillation light emission. Hence, we model the expected
voltage trace as a convolution of the impulse response function of the PMT and
the PDF model for the emission of scintillation photons:
\begin{eqnarray}
< V(t) > \propto  \int_{s = t_* }^{\infty} h_V(t-s)  f(s-t_*) ds ,
\end{eqnarray}
where $h_V(t)$ is the impulse response function of the PMT and $t_*$ is an
additional model parameter that relates the
energy deposit time $t_0$ to the relative time scale we associate with our
measurement of $h_V(t)$. We estimate the impulse response function 
by averaging single photoelectron
events observed in our calibration data. 
Neglecting additive noise and other instrumental systematic errors,
the integral of the voltage trace is proportional to the number of photoelectrons produced by the event.

The data are divided into 15 non-uniform bins by photoelectron
number, with the smallest signal bin including events consisting of 20--24
photoelectrons and the largest bin including events consisting of 240--279
photoelectrons. These bins define the region of interest. For each photoelectron
bin, we generate template traces for both electronic and nuclear recoils
by averaging all ``tagged'' events of a given type. The traces are averaged in the same manner as described in Sec.~\ref{sec:det_calibration} and normalized. 

\begin{figure}[ht]
  \centerline{
    \hbox{\psfig{figure=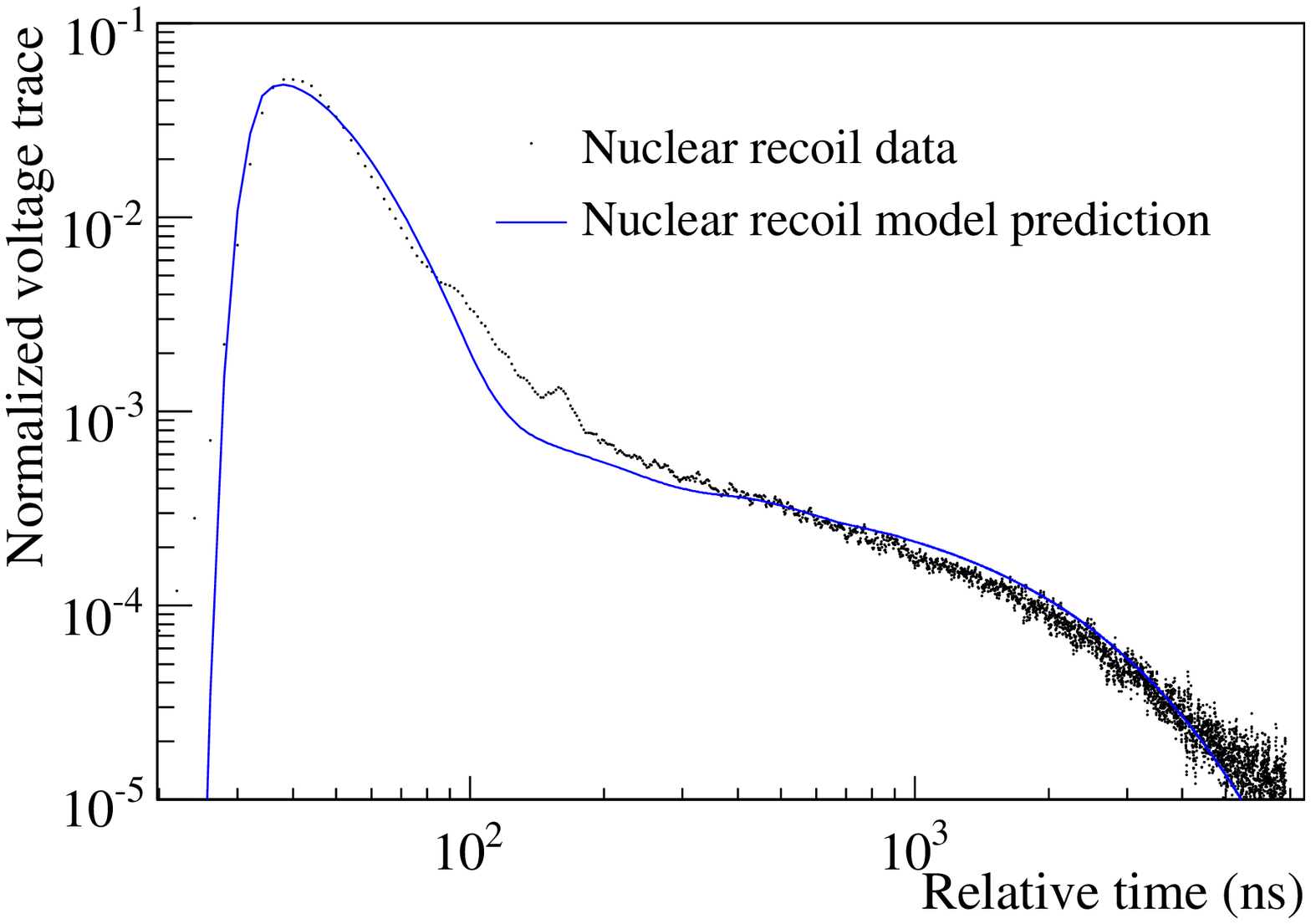,width=8cm,%
        clip=}}}
  \centerline{
    \hbox{\psfig{figure=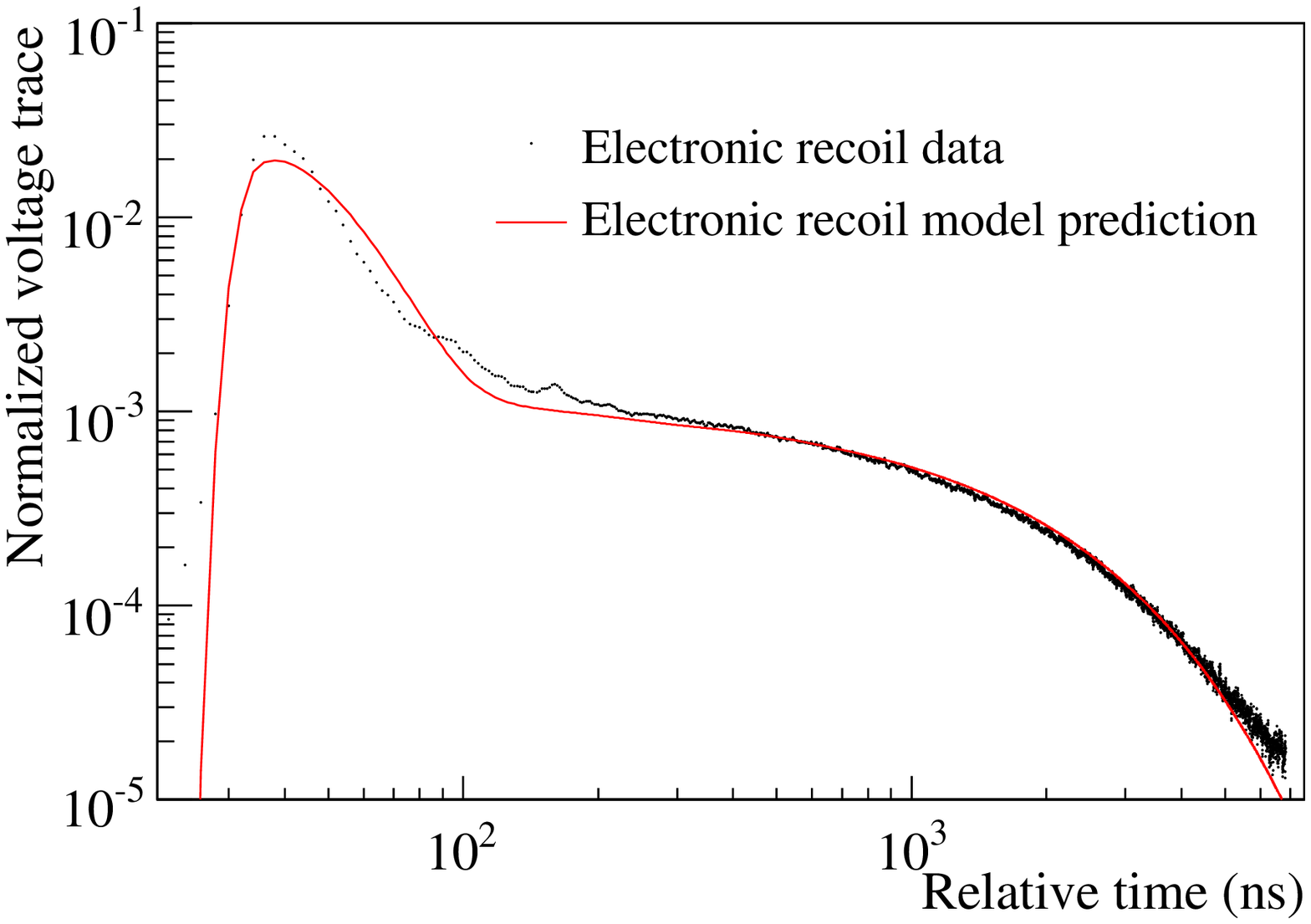,width=8cm,%
        clip=}}}
  \caption{(Color online) Observed and predicted mean voltage traces for nuclear and
    electronic recoil events of 80 to 99 photoelectrons.}
  \label{fig:fittrace}
\end{figure}

We use these average voltage traces to
determine the model parameters: $\tau_l$, $\tau_s$, $q_{nuclear}$ and $q_{electronic}$. The observed and predicted fraction of a normalized trace in the $i$th
time bin are called $p_m(i)$ and $\hat{p}(i)$, respectively. We obtain the model
parameters by minimizing the squared Matusita distance
\cite{Matusita:1954,Dillon:1978,Goldstein:1978} between $p_m$ and
$p$:
\begin{equation}
|p - p_m|_M = \sum_{i} (\sqrt{p_m(i)} - \sqrt{\hat{p}(i)})^2,
\label{Eq:matusita}
\end{equation}
where
negative 
values of $p_m$ are set to 0. For each
bin, we fit our model to normalized mean voltage trace
data in a time window that ends about 6800 ns after the
trigger time. We determine $t_*$ to be 30 ns before the trigger time, chosen to
minimize the value of Eq.~\ref{Eq:matusita}. Table~\ref{table:PDF} shows the
estimated model parameters for each photoelectron bin. Figure~\ref{fig:fittrace} shows the model prediction along with data for
the 80--99 photoelectron bin.

Between 80 and 300 ns, there is a feature in both event classes that is
not well predicted by the model. A scintillation component that decays as
approximately
$t^{-1}$ has been observed in liquid helium, attributed to
diffusion-dominated
excimer-excimer destruction~\cite{McKinsey:2003}. When we fit a model including the full form of
this component as described in~\cite{McKinsey:2003}, the parameter estimates
are not well determined.
 Previous
observations of scintillation in argon have observed an intermediate
exponential component
with a decay time of 20--40 ns~\cite{Hitachi:1983}. Unfortunately,
a model that includes a third exponential component neither returns stable
model parameters as a function of energy nor accurately predicts the trace
behavior between 80 and 300 ns, so we prefer the two-component model.  We
assume that the sharp bump localized at 150 ns is caused by the
cabling and electronics.

\begin{table}[ht]
\centering
\begin{tabular}{ccccc}\hline\hline
Bin (pe) & $\tau_l$ (ns)  &  $ \tau_s $ (ns)  & $q_{nuclear} $  & $q_{electronic}$ \\ \hline
   20--24& $1634\pm150$&$  9\pm  3$&$     0.378\pm     0.011$&$     0.523\pm     0.015$\\
   25--29& $1535\pm128$&$  10\pm  3$&$     0.382\pm     0.011$&$     0.573\pm     0.015$\\
   30--34&$1478\pm107$&$ 10\pm  3$&$     0.357\pm     0.010$&$     0.601\pm     0.014$\\
   35--39&$1455\pm102$&$ 11\pm  3$&$     0.353\pm     0.010$&$     0.627\pm     0.014$\\
   40--49&$1461\pm    96$&$ 12\pm  3$&$     0.344\pm     0.010$&$     0.658\pm     0.014$\\
   50--59& $1459\pm    92$&$ 12\pm  3$&$     0.327\pm     0.009$&$     0.681\pm     0.015$\\
   60--69&$1439\pm    89$&$ 12\pm  3$&$     0.315\pm     0.010$&$     0.699\pm     0.015$\\
   70--79&$1448\pm    89$&$ 13\pm  3$&$     0.309\pm     0.010$&$     0.710\pm     0.015$\\
   80--99& $1447\pm    85$&$ 13\pm  3$&$     0.298\pm     0.010$&$     0.721\pm     0.015$\\
  100--119& $1452\pm    84$&$ 13\pm  3$&$     0.289\pm     0.010$&$ 0.733\pm    0.015$\\
  120--139&$1447\pm    84$& $13\pm  3$&$     0.284\pm     0.011$&$     0.741\pm     0.016$\\
  140--159&$1446\pm    84$&$ 14\pm  3$&$     0.278\pm     0.012$&$     0.747\pm     0.016$\\
  160--199&$1450\pm    84$&$ 14\pm  3$&$     0.272\pm     0.013$&$     0.752\pm     0.016$\\
  200--239& $1460\pm    84$&$ 15\pm  3$&$     0.265\pm     0.015$&$     0.760\pm     0.016$\\
  240--279&$1467\pm    84$&$ 15\pm  3$&$     0.258\pm     0.018$&$
  0.764\pm     0.016$\\ \hline\hline

\end{tabular}
\caption{Estimated model parameters
and
1-sigma uncertainties 
for each photoelectron bin. The systematic errors described in the text are
the dominant source  of error. }
\label{table:PDF}
\end{table}

We determine 1-sigma
random uncertainties for the model parameters with a
nonparametric bootstrap resampling scheme~\cite{Efron:1993}. 
A rigorous quantification of systematic uncertainty arising from the fit
window, the voltage applied to the PMT, and variation in the response of each
individual PMT
is difficult.
Nonetheless, we get approximate estimates of
systematic uncertainties from these sources, and these systematics are the
dominant source of error.
We estimate
1-sigma systematic uncertainties in the parameters $\tau_i$ (for $i=l,s$)
as
$0.5\times (\tau_{i,max} - \tau_{i,min})$,
where $\tau_{i,max}$ and $\tau_{i,min}$ are the maximum and minimum values
taken by $\tau_i$ for different choices of time windows.
We
refit the model to data using time windows
of approximately 5000 ns and 8600 ns from the trigger time,
and
vary the offset parameter $t_{*}$ by $\pm $ 2 ns. We also check for
differences between the response of each PMT in five different photoelectron bins. We include the voltage-dependent errors described in
Sec.~\ref{sec:det_calibration} in the systematic uncertainty of
$\tau_l$, and we estimate the voltage-dependent systematic variation of the $q$-values.  We find that the time constants are most influenced by
choice of fit window, while the $q$-values are most influenced by PMT
effects. We combine all sources of error to obtain the estimates of systematic
uncertainty shown in Table~\ref{table:PDF}.

As a consistency check on the quality of fit, for each photoelectron bin we
use our model to predict the prompt fraction, a discrimination statistic
described in detail in the next section. We compute the difference between the predicted and observed prompt
fractions, and the root-mean-square values of this prediction error across
all bins are respectively 0.007 and 0.003 for the nuclear and electronic recoil event
classes. The fractional root-mean-square value of this prediction error is about
$1\%$ for both event classes.

  \subsection{Prompt Fraction Method}
  \label{sec:PromptFraction}

\begin{figure}[ht]
  \centerline{
    \hbox{\psfig{figure=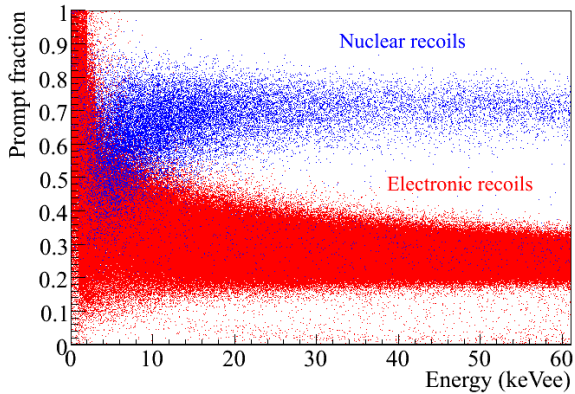,width=8cm,%
        clip=}}}
  \caption{(Color online) A scatter plot of \fP\, vs. energy for tagged electronic and nuclear
    recoils, where $\xi = 90$ ns.}
  \label{fig:scatter}
\end{figure}

The prompt fraction method is a simple approach to pulse shape
discrimination. For each trace, we define the prompt fraction \fP\, as
\begin{equation}
  f_{p} = \frac{\int_{T_i}^{\xi} V(t) dt}{\int_{T_i}^{T_f} V(t) dt},
  \label{eq:fp}
\end{equation}
where $V(t)$ is the voltage trace from the PMT, $\xi$ is an
integration time determined to optimize the ERC, $T_i = t_0 - 50$ ns, $T_f =
t_0 + 9$ \mus, and $t_0$ is the trigger time as defined in
Sec.~\ref{sec:data_collection}. The measured discrimination does not significantly improve by extending
$T_f$ to 20 \mus. Fig.~\ref{fig:scatter}
shows a scatter plot of \fP\, versus energy for both electronic and nuclear
recoils. The two populations of events represent the
tagged data remaining in the neutron generator data set and the \Na\, data set after the selection cuts described in
Sec.~\ref{sec:data_collection} have been made. We choose $\xi = 90$~ns by
estimating the ERC based on a simple Gaussian model for values of $\xi$ from 50 to 250~ns over a variety of
different photoelectron bins. 
Although the value of $\xi$ has only a weak effect on the
predicted discrimination, a choice of $\xi = 90$ ns provides the
best results across the widest range of energies in the region of
interest, and that value is used for the analysis presented in this paper.

\begin{figure}[ht]
  \centerline{
    \hbox{\psfig{figure=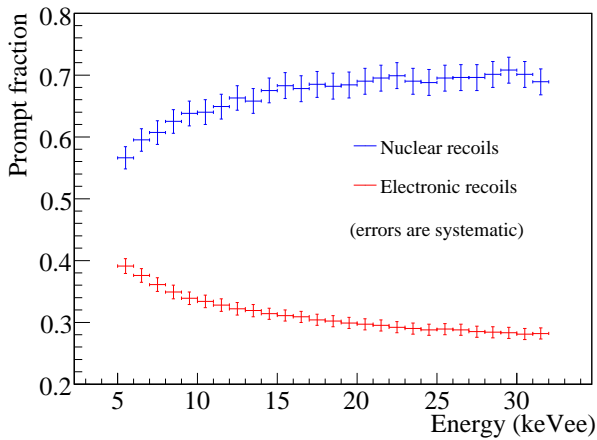,width=8.6cm,%
        clip=}}}
  \caption{(Color online) The estimated mean \fP\, versus energy for both
    event classes where $\xi =
    90$ ns.} 
  \label{fig:fpvenergy}
\end{figure}

For energy bins of width 1 keVee between 5 and 32 keVee, we form histograms of the electronic and
nuclear recoil \fP\, statistics. To estimate the expected value of $f_p$, $\hat{f_p}$, we fit a
Gaussian function to the empirical distributions. 
In
Table~\ref{table:fp} and Fig.~\ref{fig:fpvenergy}, we present the estimated mean \fP\, for both classes of events in
the energy range of interest. We estimate the systematic uncertainty on the values in Table~\ref{table:fp} to be $3\%$. This uncertainty estimate comes
from changing the PMT voltage by $-75$ V and from variations in the measured signals between the
two PMTs. The mean values of the \fP\, distributions for the two event classes are closer at low
energies than at high energies, possibly because $dE/dx$ for nuclear recoils decreases at low
energies while increasing for electronic recoils. Therefore, the PSD improves
at higher energies both because of increased photoelectron
statistics and because of increased separation between the mean \fP\, values.

\begin{table}[ht]
\centering
\begin{tabular}{ccc}\hline\hline
Energy (keVee) & $\hat{f}_{p,electronic}$ & $\hat{f}_{p,nuclear}$ \\\hline
5--6 & $0.391\pm0.012$ &  $0.566\pm0.018$ \\ 
6--7 & $0.376\pm0.011$ & $0.595\pm0.018$ \\
7--8 & $0.361\pm0.011$ & $0.607\pm0.019$ \\
8--9 & $0.349\pm0.011$ & $0.625\pm0.019$ \\
9--10 & $0.339\pm0.010$ & $0.638\pm0.020$ \\
10--11 & $0.334\pm0.010$ & $0.640\pm0.020$ \\
11--12 & $0.328\pm0.010$ & $0.649\pm0.020$ \\
12--13 & $0.322\pm0.010$ &  $0.663\pm0.020$ \\
13--14 & $0.319\pm0.010$ & $0.658\pm0.020$ \\
14--15 & $0.314\pm0.009$ &  $0.675\pm0.020$ \\
15--16 & $0.311\pm0.009$ & $0.683\pm0.021$ \\
16--17 & $0.309\pm0.009$ & $0.678\pm0.021$ \\
17--18 & $0.304\pm0.009$ & $0.685\pm0.021$ \\
18--19 & $0.302\pm0.009$ & $0.682\pm0.021$ \\
19--20 & $0.299\pm0.009$ & $0.684\pm0.021$ \\
20--21 & $0.297\pm0.009$ & $0.690\pm0.021$ \\
21--22 & $0.295\pm0.009$ & $0.695\pm0.021$ \\
22--23 & $0.292\pm0.009$ & $0.699\pm0.021$ \\
23--24 & $0.290\pm0.009$ & $0.690\pm0.021$ \\
24--25 & $0.288\pm0.009$ & $0.688\pm0.021$ \\
25--26 & $0.289\pm0.009$ & $0.695\pm0.021$ \\
26--27 & $0.288\pm0.009$ & $0.696\pm0.021$ \\
27--28 & $0.285\pm0.009$ & $0.696\pm0.021$ \\
28--29 & $0.284\pm0.009$ & $0.701\pm0.021$ \\
29--30 & $0.283\pm0.009$ & $0.708\pm0.021$ \\
30--31 & $0.281\pm0.009$ & $0.701\pm0.021$ \\
31--32 & $0.282\pm0.009$ & $0.689\pm0.021$ \\\hline\hline

\end{tabular}
\caption{This table presents estimated mean values of  \fP\, versus energy,
  where $\xi = 90$ ns. The main sources of uncertainty are systematic,
  stemming from voltage effects and differences in the measured signals between the two PMTs.}
\label{table:fp}
\end{table}

We estimate the ERC as the
number of tagged electronic recoil events with $f_{p}>
\hat{f}_{p,\mathrm{nuclear}}$ divided by the total number of electronic recoil events, where $\hat{f}_{p,\mathrm{nuclear}}$ is the estimated mean \fP\, for
nuclear recoils of that energy. This restriction
sets a nuclear recoil acceptance level of approximately 50\%. Since the
shielding, coincidence, and timing cuts do not eliminate all neutron
backgrounds in the detector, a background estimation $N_{\mathrm{bg}}$ is made by measuring the
rate of background neutrons $R_{\mathrm{bgn}}$ and assuming that the background is
dominated by these neutrons hitting in accidental coincidence with $\gamma$ rays ($R_{\gamma})$ in
the liquid scintillator during the time allowed by the time-of-flight cut:
\begin{equation}
N_{\mathrm{bg}} = R_{\mathrm{bgn}} \times R_{\gamma} \times \mathrm{TOF} \times{T_{a}},
\end{equation}
where $T_a$ is the acquisition time of the data. To enable comparison with the measured
ERC, we divide $N_{\mathrm{bg}}$ for each energy bin by the total number of electronic
recoil events in that bin.

\begin{figure}[ht]
  \centerline{
    \hbox{\psfig{figure=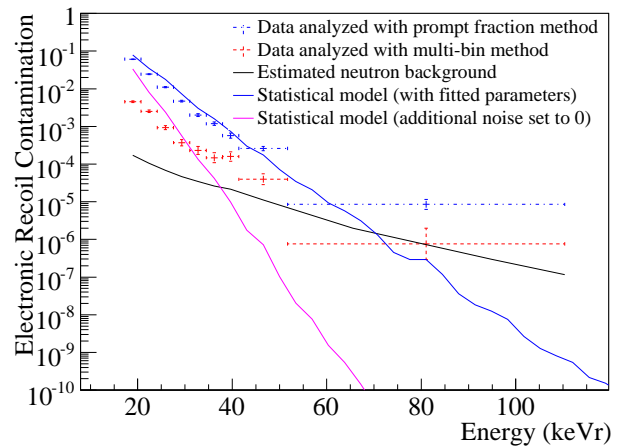,width=8.6cm,%
        clip=}}}
  \caption{(Color online) Measured electronic recoil contamination obtained by using the prompt fraction
    method and the multi-bin method. Also shown: the background estimation and model predictions
    described in the text. We include a model with the additional noise set to 0
    for comparison. There were no contamination events above 69
    keVr observed by use of either method. 
    The energy axis has been scaled from 
    keVee to keVr by use of a constant nuclear recoil scintillation efficiency
    of 0.29 as discussed in the text.} 
  \label{fig:AA_57}
\end{figure}

Figure~\ref{fig:AA_57} shows the ERC observed
using the prompt fraction method. We also plot the background estimation, two
PSD projections based on the
statistical model described below, and the ERC observed by applying a multibin method
of PSD described in the next section. We convert the energy axis in
Fig.~\ref{fig:AA_57} to keV of nuclear recoil energy (keVr) from keVee by
dividing all electron equivalent energies by a constant nuclear recoil
scintillation efficiency of 0.29. This value was obtained by measurements
using the same apparatus described in this paper and will be discussed in an
upcoming publication~\cite{Gastler:2008}. We present the PSD results in keVr because that is the
unit of interest for a dark matter detector.  Using the prompt fraction
method, for a nuclear recoil acceptance level of approximately $50\%$, we measure a background- and statistics-limited level of
ERC in our detector of $8.5\times10^{-6}$ between 52 and 110 keVr (11
contamination events). We observe
no nuclear recoil-like events above 69 keVr. For comparison, there is an uncorrelated neutron background
rate of $\sim 6$ mHz between 69 and 110 keVr, corresponding to 0.25 expected
background counts in that energy range.

Following work done by members of the DEAP
collaboration~\cite{Lidgard:2008,Boulay:2008}, we model our estimates of the number of photoelectrons in the prompt and late
time windows, $N_p$ and $N_l$, as normally distributed, independent random
variables with means $\mu_p$ and $\mu_l$ and variances $\sigma_p^2$ and
$\sigma_l^2$. The estimated total number of photoelectrons, $N_{\mathrm{tot}} = N_p + N_l$, is also
a random variable, with mean $\mu_{\mathrm{tot}} = \mu_p + \mu_l$ and variance
$\sigma^2_{\mathrm{tot}} = \sigma_p^2 + \sigma_l^2$. We express $\mu_p$
and $\mu_l$ in terms of $\mu_{\mathrm{tot}}$ and $\hat{f_p}$ and we decompose the variances
into two components:  
\begin{eqnarray}
\nonumber \mu_p& =& \hat{f_{p}}\mu_{\mathrm{tot}}\\
\nonumber \mu_l& =& (1 - \hat{f_{p}})\mu_{\mathrm{tot}}\\
\nonumber \sigma_p^2& =& \mu_p + {\sigma^2_{p,\mathrm{add}}}\\
 \sigma_l^2& =& \mu_l + {\sigma^2_{l,\mathrm{add}}}
\end{eqnarray}
where $\sigma_{p,\mathrm{add}}$ and $\sigma_{l,\mathrm{add}}$ represent additional sources of random
variability beyond what we expect from Poisson counting statistics (for example, integration
noise). 

Hinkley~\cite{Hinkley:1969} has described in detail the probability
density function of the ratio of two normally distributed, correlated random
variables. For simplicity, we present here an approximation [given by Eq.~9 of
Ref.~\cite{Hinkley:1969}] to the PDF of $f_p = N_p/N_{\mathrm{tot}}$:
\begin{eqnarray}
\nonumber g_{f_p}(x) =& \frac{\sigma_l^2\mu_p x+\sigma_p^2\mu_l (1 -
x)}{\sqrt{2\pi}(\sigma_l^2 x^2+\sigma_p^2 (1 -
x)^2)^{3/2}}\, \times \\
&\mathrm{exp}[-\frac{(\mu_l x-\mu_p (1 -
x)^2)}{2(\sigma_l^2 x^2+\sigma_p^2 (1 -x)^2)}],
\label{eqn:MF}
\end{eqnarray}
where we have used the fact that the correlation, $\rho$, between ${N_p}$ and ${N_{\mathrm{tot}}}$ is
\begin{equation}
\rho = \frac{\sigma_p}{\sqrt{\sigma_p^2+\sigma_l^2}}.
\end{equation}
For the analysis below, we use the exact PDF [Eq.~1 in Ref.~\cite{Hinkley:1969}], although in practice the approximation is
extremely good down to the lowest energy bin examined.   

We fit the electronic recoil data
with the statistical model in each energy bin by fixing $N_{\mathrm{tot}}$ according to our measured light yield
and treating $\hat{f_p}$, $\sigma_{p,\mathrm{add}}$, and $\sigma_{l,\mathrm{add}}$ as free
parameters. We
assume that the statistical distribution of $f_p$ does not strongly depend on
the value of $N_{\mathrm{tot}}$ for events of the same energy. For any particular
energy deposit, $N_{\mathrm{tot}}$ is a random variable, and events due to many
different energy deposits can contribute to prompt ratio data in any one bin
in $N_{\mathrm{tot}}$ space. The probability density function for prompt ratio data is
a mixture of energy dependent PDFs, and we neglect this energy
blurring effect. Simple Monte Carlo
studies suggest that the ratio-of-Gaussians model breaks down for idealized
Gaussian data due to the constraint
on $N_{\mathrm{tot}}$ resulting from the binning of data, and this effect has not been
taken into account in our analysis. 

 We use the statistical model to estimate the expected fraction of electronic recoils that are misclassified as nuclear recoils:
\begin{equation}
  \mathrm{ERC} = \int_{\eta}^1 g_{f_{p}}(x) dx.
  \label{eq:ERA}
\end{equation}
Here, we choose $\eta = \hat{f}_{p,\mathrm{nuclear}}$ to set the nuclear recoil
acceptance level to approximately 50\%, and we choose the parameters of
$g_{f_{p}}$ according to the fits to the electronic recoil \fP\,
distribution. Fig.~\ref{fig:AA_57} shows the predicted ERC versus energy
according to this model. We also plot the idealized case where $\sigma_{l,\mathrm{add}}$ and
$\sigma_{p,\mathrm{add}}$ are set to 0. In general, we expect that the normal
distribution model for $N_p$ and $N_l$ is an approximation for any
energy of interest.  For the idealized case where $\sigma_{p,\mathrm{add}} =
\sigma_{l,\mathrm{add}} = 0$, $N_p$ and $N_l$ would be Poisson random variables
rather than Gaussian random variables.  Hence, the ERC  predicted
by the ratio-of-Gaussians model for the idealized case should be
interpreted with caution particularly at lower energies where the
accuracy of a normal distribution model for a Poisson random variable
can be very poor.

\begin{figure}[ht]
  \centerline{
    \hbox{\psfig{figure=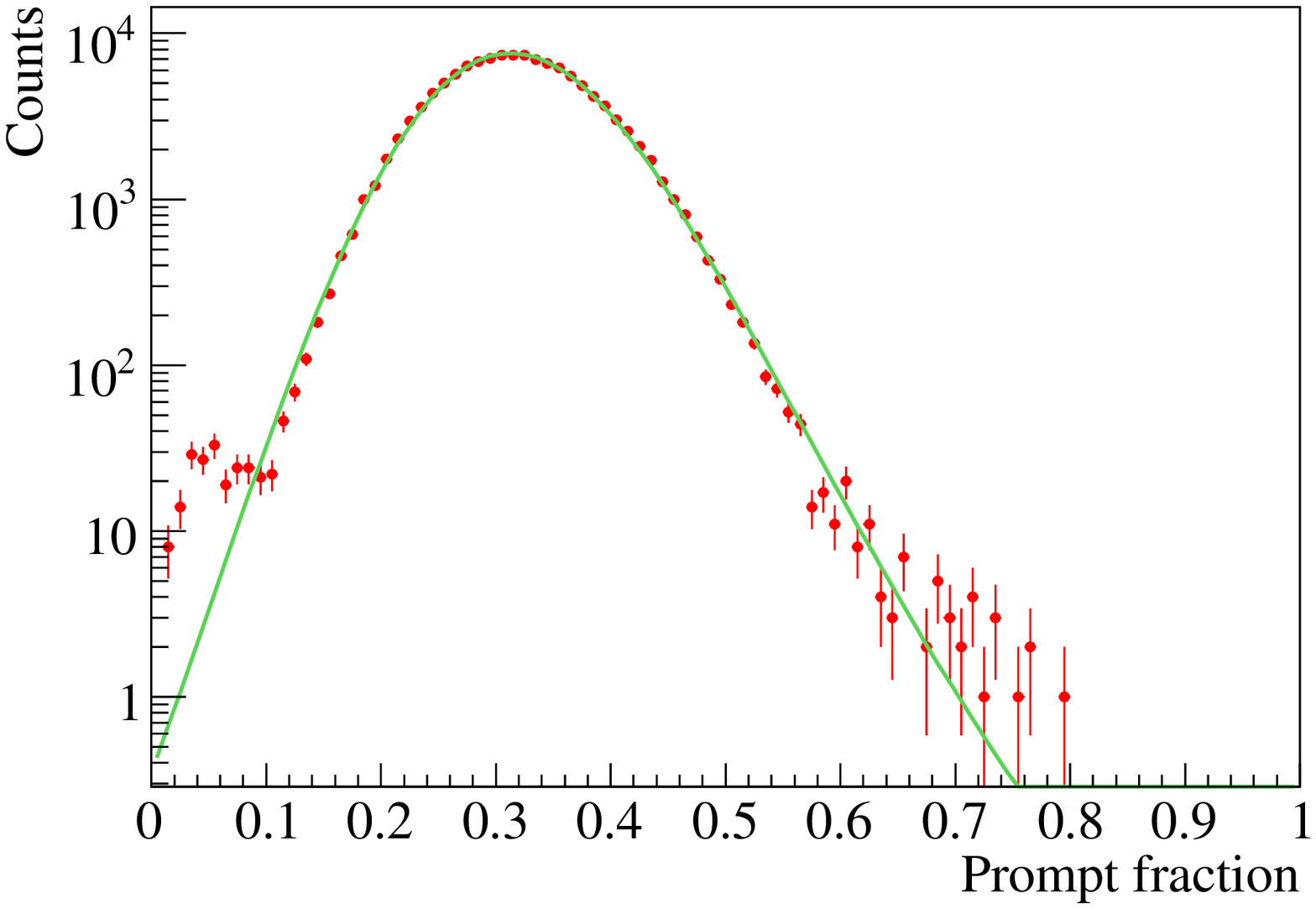,width=8cm,%
        clip=}}}
  \centerline{
    \hbox{\psfig{figure=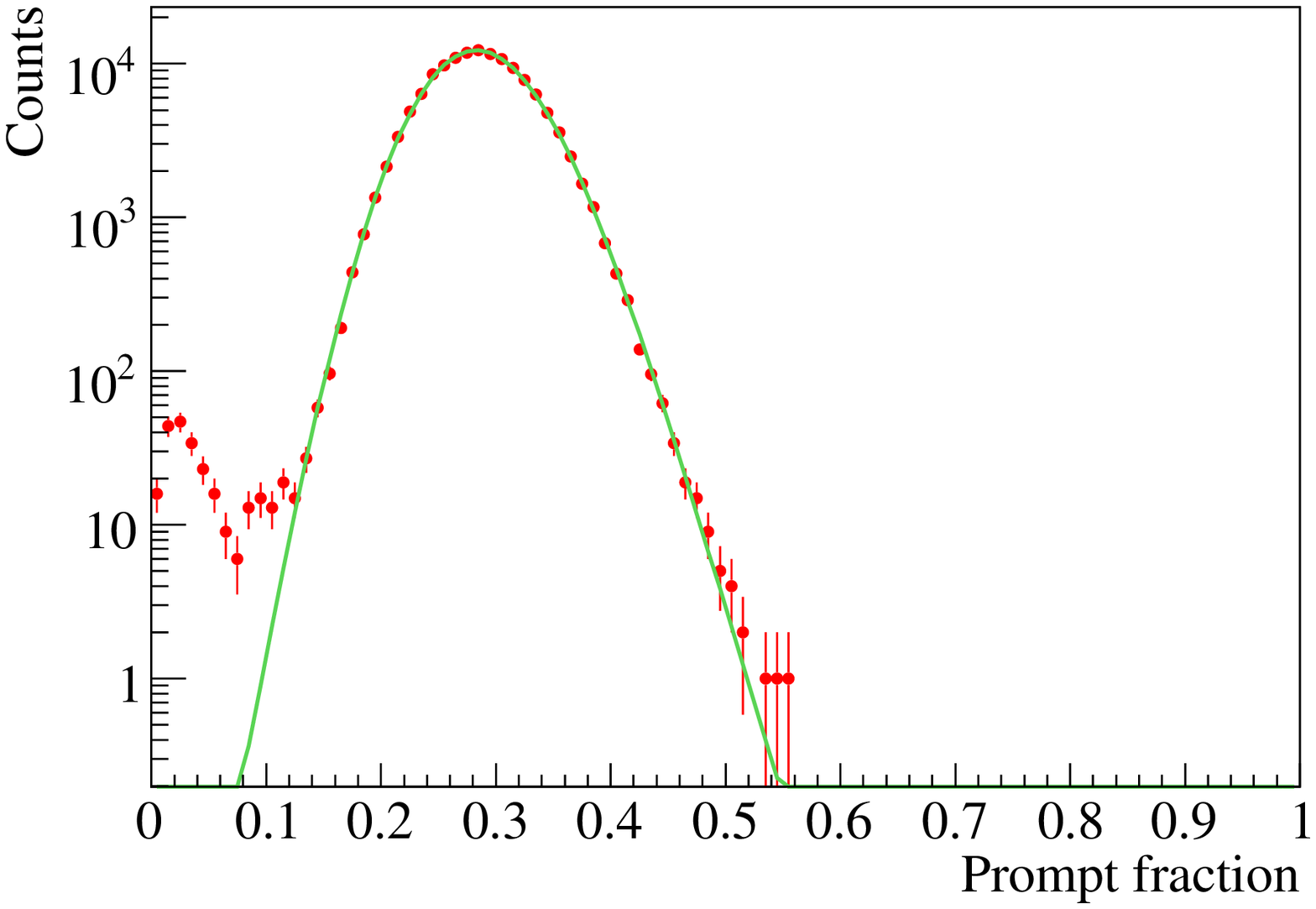,width=8cm,%
        clip=}}}
  
  \caption{(Color online) Projections of the electronic recoil data from Fig.~\ref{fig:scatter} onto the y-axis for 14--15 keVee
   (top) and 30--31 keVee events (bottom), fitted by Eq.~\ref{eqn:MF} as discussed
     in the text.}
  \label{fig:statmodel}
\end{figure}

Figure~\ref{fig:statmodel} shows an example of the model fit for
14--15 keVee and 30--31 keVee electronic recoil events. There is a deviation from the model at low \fP\, values that we
attribute to pile-up and noise triggers. There is also an
excess of events in the high \fP\, region. This excess might be caused by edge
effects 
in our detector coming from $\gamma$ tracks going into the walls or the TPB
layer,
producing extra prompt light. A larger detector with position reconstruction capability might be able
to eliminate such edge effects. There could be some unknown
phenomena at work in the production and decay of argon molecular states,
yielding a small fraction of events with anomalously large \fP\, values. A third possibility is approximation error stemming from incorrect
model assumptions, such as the Gaussianity of $N_p$ and $N_l$ or the effect
of
data binning.  These
possible effects can be better investigated by detectors with improved neutron
shielding, and future studies will require more work to better calibrate the
ERC predictions made by the
 ratio-of-Gaussians model.

\begin{figure}[ht]
  \centerline{
    \hbox{\psfig{figure=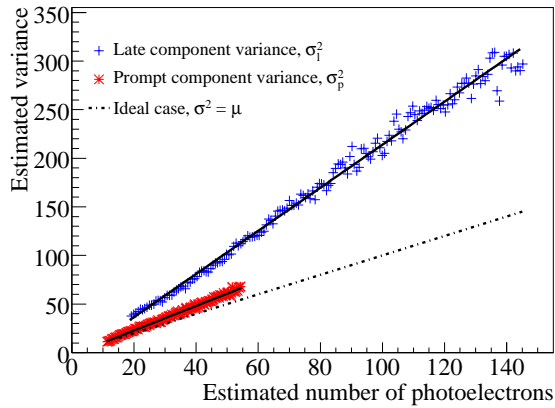,width=8cm,%
        clip=}}}
    \caption{(Color online) Estimated $\sigma_{l}^2$ and $\sigma_{p}^2$ parameters from the
      statistical model plotted against the estimated number of photoelectrons in the
      late and prompt components, respectively. Each distribution is fitted by a straight
      line, and the ideal case where $\sigma^2 = \mu$ is also shown.  }
  \label{fig:noise}
\end{figure}

Figure~\ref{fig:noise} shows the fitted values of the variances,
$\sigma_{l}^2 = \mu_l + \sigma^2_{l,\mathrm{add}}$ and $\sigma_{p}^2 = \mu_p + \sigma^2_{p,\mathrm{add}}$ versus the
 estimated number of photoelectrons in $\mu_l$ and $\mu_p$.  The ideal case
of zero additional noise ($\sigma^2 = \mu$) has been shown for comparison.  The relationship
 between the estimated variance parameters and the corresponding estimates of
 the number of photoelectrons is approximately linear, suggesting that
 $\sigma^2_{l,\mathrm{add}}$ and $\sigma^2_{p,\mathrm{add}}$ are proportional to $\mu_l$ and
 $\mu_p$.  Straight line fits to the late and prompt variances have slopes of 2.2 and 1.3,
respectively. Due to the various uncertainties in the fit such
as that associated with data binning, these numbers do not represent a rigorous estimate of the amount of noise in the
late and prompt distributions; however, they might be used as a point of comparison
with future detectors in trying to reduce the overall noise in the system. Possible sources of this additional noise include integration
noise, the widths of the
single photoelectron spectra, and
variability in the prompt window size arising from uncertainty in the determined
trigger position. To improve PSD, future experiments will need to
reduce the size of this additional noise to approach the ideal case.

  \subsection{Multibin Method}
  \label{sec:ModMaxLike}

The prompt fraction method of PSD is based on binning the voltage trace into
two time bins. We generalize this approach
by representing a normalized voltage trace as a $K\times L$ dimensional matrix. In
this representation, we categorize the data by the number of
photoelectrons in the event, and $L$ refers to this division of data into
photoelectron bins. We also
partition the voltage trace into $K$ time bins. We
choose $K=10$ for ease of computation, but there may be a better choice
of $K$. With the exception of the four smallest signal bins, which have been
combined to form bins of 20--29 and 30--39 photoelectrons, we use the same average voltage traces obtained in Sec.~\ref{sec:TimePDF}
as templates.

We partition each template into $K=10$
time bins.  The upper and lower endpoints of each time bin are selected
so that the fractions of the template trace for 80--99 photoelectron electronic
recoil events that falls in each time bin are
approximately equal. The initial bin starts 50 ns before the trigger, and the
endpoints of each bin are as follows, measured in nanoseconds from the trigger: 8, 18,
56, 200, 440, 750, 1180, 1800, 2950, and 8000. We do not adjust these endpoints for different
photoelectron number but
use the 80--99 photoelectron based time binning scheme for all cases.

The $k$th component of the normalized template
for an event that
falls in the
$l$th photoelectron bin
for the nuclear and electronic recoil classes
is denoted $p_n(k,l)$ or $p_e(k,l)$, respectively. For instance, $p_n(1,1)$ represents the
fraction of the 20--29 photoelectron nuclear recoil template trace that falls
in the first time bin (between 50 ns
before the trigger and 8 ns after the trigger). 

To assign an event to either the nuclear recoil or electronic
recoil class,
we first compute a discrimination statistic for the event.
We motivate a discrimination statistic based on
analysis of an
idealized experiment in which we could observe the absolute detection time of each photoelectron
without error.
In this ideal case, for a fixed number of detected photoelectrons, 
the observed number of photoelectrons in the time bins
is a multinomial 
random variable.

Given that the
fraction of detected photoelectrons in the $k$th bin is
$p_m(k,l)$,
the multinomial log-likelihood statistics for the nuclear and electronic recoil classes are
\begin{equation}
\ln {Y_n} = N_{\mathrm{tot}}\sum_{k=1}^{K} \sum_{l=1}^{L}  \delta_{ll'} p_m(k,l) \ln{ p_n(k,l')
 } + \mathrm{const}
\label{Eq:logn}
\end{equation}
and
\begin{equation}
\ln { Y_e}= N_{\mathrm{tot}} \sum_{k=1}^{K} \sum_{l=1}^{L}  \delta_{ll'} p_m(k,l) \ln{
  p_e(k,l') } + \mathrm{const},
\label{Eq:loge}
\end{equation}
respectively~\cite{Mood:1974}. Here,
$N_{\mathrm{tot}}$ is the total number of detected photoelectrons,
and
$p_n(k,l)$ and 
$p_e(k,l)$ are the expected values of
the fraction of detected photoelectrons in the
$k$th time bin
for the nuclear and electronic recoil classes, respectively.
For 
this idealized case,
a natural choice for the discrimination statistic is
the log-likelihood ratio
statistic, $\ln{R_m}$:

\begin{equation}
 \ln { R_m } = \ln{ Y_e } - \ln{ Y_n}.
\label{Eq:logratio}
\end{equation}

In our experiment, the observed data is not
multinomial, primarily because
we observe a noisy voltage waveform rather than 
discrete detection times.
Nonetheless, we compute a discrimination statistic using
Eqns.~\ref{Eq:logn}--\ref{Eq:logratio}
where we estimate $N_{\mathrm{tot}}$
and $p_m(k,l)$
from any voltage trace of interest,
and determine
$p_e(k,l)$ and $p_n(k,l)$ 
as described earlier.

\begin{figure}[ht]
  \centerline{
    \hbox{\psfig{figure=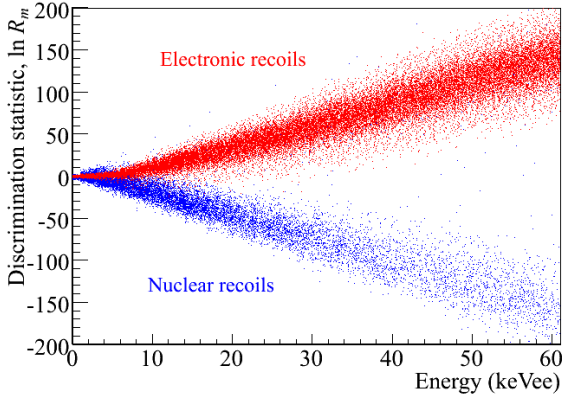,width=8cm,%
        clip=}}}
  \caption{(Color online) A scatter plot of $\ln R_m$ vs. energy for both electronic and nuclear
     recoils.}
  \label{fig:ratio}
\end{figure}

\begin{figure}[ht]
  \centerline{
    \hbox{\psfig{figure=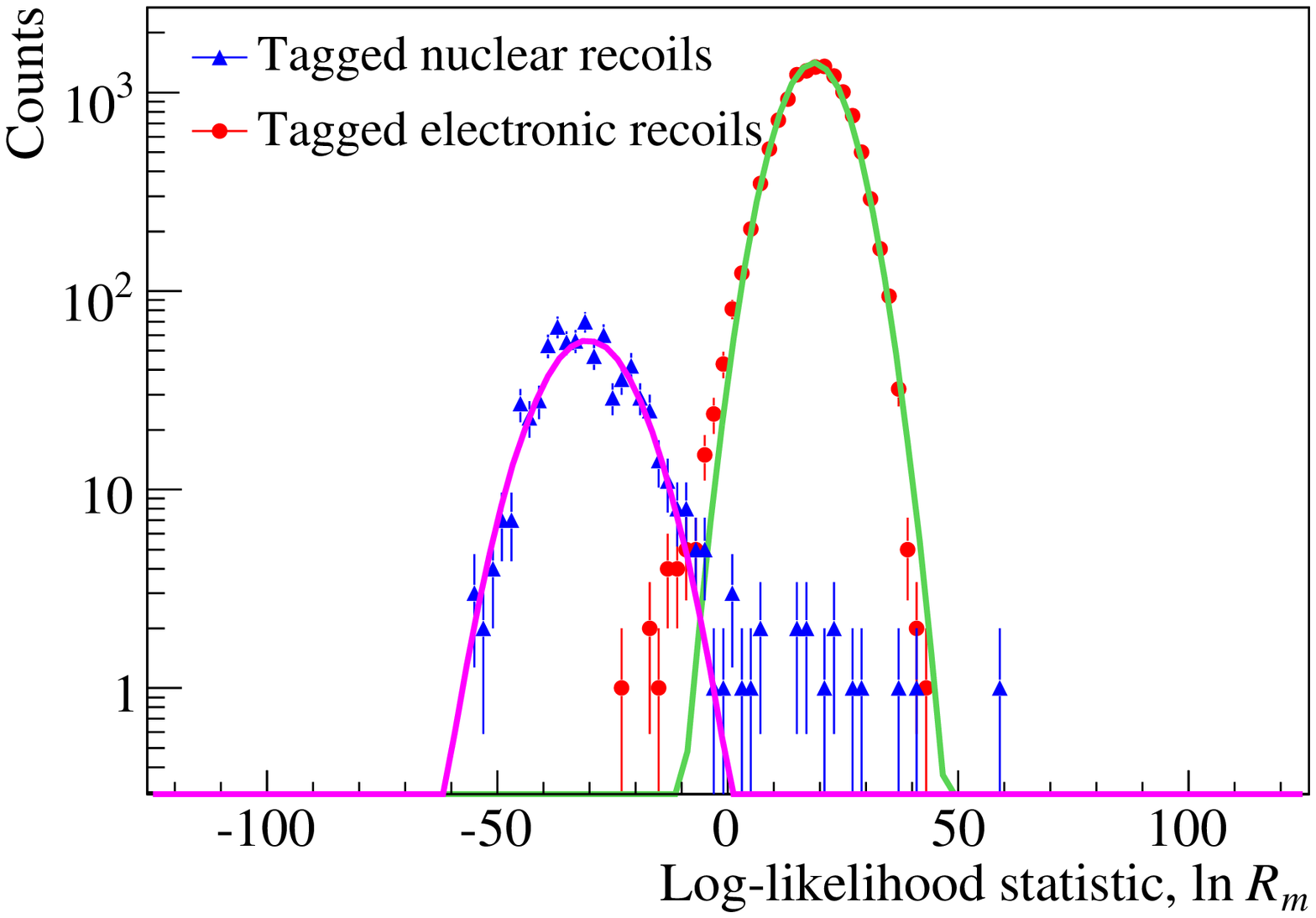,width=8cm,%
        clip=}}}  \centerline{
    \hbox{\psfig{figure=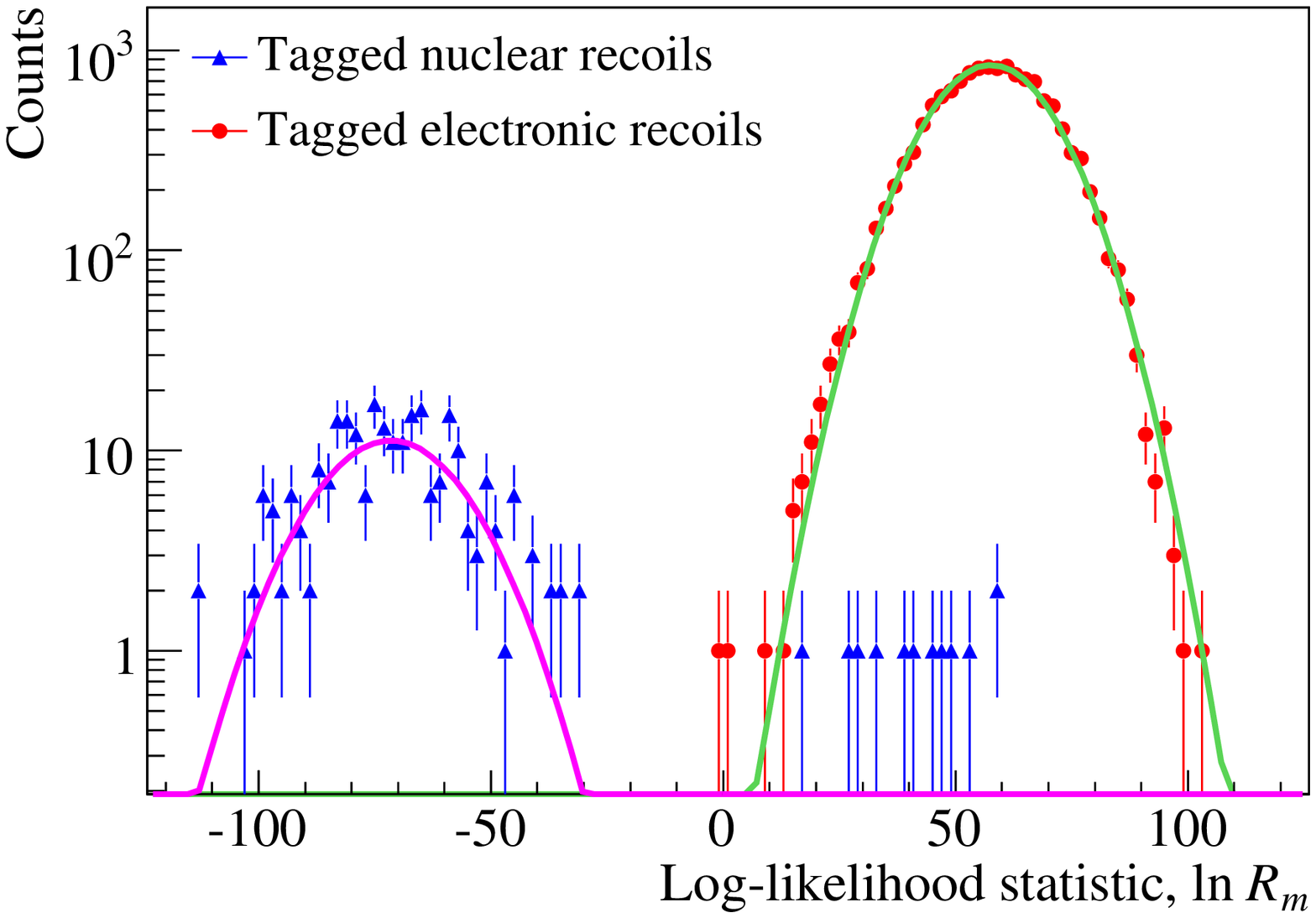,width=8cm,%
        clip=}}}
  \caption{(Color online) Projections of Fig.~\ref{fig:ratio} onto the y-axis for 14--15
    keVee (top) and 30--31
    keVee (bottom) events, with Gaussian fits to both the electronic (left) and nuclear
    (right) recoil distributions.}
  \label{fig:maxgaussian}
\end{figure}

Figure~\ref{fig:ratio} shows a scatter
plot of $\ln {R_m}$ versus energy, analogous to Fig.~\ref{fig:scatter}. From this point the analysis
parallels the prompt fraction method, as we form histograms of $\ln {R_m}$ by energy bin and fit a Gaussian function 
to the observed $\ln {R_m}$ statistics to estimate the mean of $\ln {R_m}$. Fig.~\ref{fig:maxgaussian}
shows an example of the fitted projections for 14--15 keVee and 30--31 keVee
events.

For each energy bin,
we estimate the mean values of $\ln{R_m}$ for nuclear recoils based on the
Gaussian fits to the observed distributions, and this estimated mean value determines an approximate $50\%$ nuclear acceptance
threshold. We then
determine the fraction of
events in the tagged electronic recoil data set that
have discrimination statistics less than this mean value to determine 
the observed level of ERC using the multi-bin method. If we
determine the $50\%$ nuclear acceptance threshold by finding the median values of the
nuclear recoil distributions, the observed ERC is not significantly affected.
Figure~\ref{fig:AA_57} in  Sec.~\ref{sec:PromptFraction} shows the ERC using
the multibin method. The multibin method outperforms the prompt fraction
method by as much as an order of magnitude. For a nuclear recoil
acceptance of approximately $50\%$, we have measured a
background- and statistics-limited ERC of
$7.6\times10^{-7}$ between 52 and 110 keVr (1 contamination event). We observe no contamination
events above 62 keVr using the multibin method.

\section{Maximum Likelihood PSD}

\label{sec:MaxLike}

We predict the PSD achievable in a large detector with many PMTs
by assuming we can measure discrete times for each photoelectron detected in
an event. We simulate events in this detector using the PDFs measured in Sec.~\ref{sec:TimePDF} and assume that
the detection times of photoelectrons are measured without error. We develop a
maximum likelihood PSD method and apply it to the simulated data.

For an event that generates $N$ photoelectrons,
we denote the
detection times as $\tb = (t_1,t_2,...,t_N)$. We define the log-likelihood
function, $\ln L$, of this data following Ref.~\cite{Cleveland:1983} as
\begin{equation}
\ln L = -m + \sum_{i=1}^{N} \ln(r_{bg} + \lambda f(t-t_0)),
\end{equation}
where
\begin{equation}
m ~=~ \int_{T_b} ^ {T_e} (r_{bg} + \lambda f(t-t_0)) dt,
\end{equation}
$t_0$ is the time at which the energy deposit occurs,
$f(t-t_0)$ is the PDF for the observed photoelectrons [Eq.~\ref{Eq:PDF}], $\lambda$ is the
expected number of detected photoelectrons generated by the event, $T_b$ and
$T_e$ are the start and end time of the observation, and $r_{bg}$ is the
background rate.

In our simulations, we set $r_{bg} = 0$ and $t_0 > T_b$. The maximum
likelihood estimates of $t_0$ and $\lambda$ are therefore $t_1$, the first detection
time, and $\hat{\lambda}$, where
\begin{equation}
\hat{\lambda} = \frac{N}{qF(T_e - t_1,\tau_1) + (1-q) F(T_e - t_1,\tau_2)}
\end{equation}
and $F(T,\tau) = 1 - \exp(-T/\tau)$.  Consequently, $m = N$. Additionally, the
time window in our simulations is 40 $\mu$s, so $\int_{T_b}^{T_e}
f(t-t_0) dt \approx 1$.  Thus, for our case the likelihood function of the
observed data is well approximated as
\begin{equation}
L(\tb ) \approx \exp(-N)\,N^N \prod_{i=1}^{N} f( t_i - t_1).
\end{equation}

\begin{figure}[ht]
  \centerline{
    \hbox{\psfig{figure=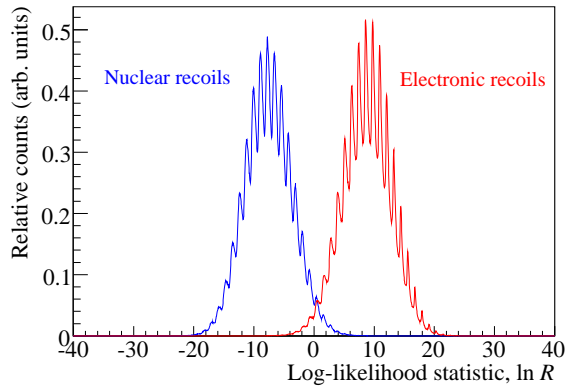,width=8cm,%
        clip=}}}
  \caption{(Color online) Monte Carlo estimates of the log-likelihood
statistics for events that yield 50 photoelectrons.}
  \label{fig:MLlogR}
\end{figure}

Following Ref.~\cite{Davies:1994} and the discussion above, for each simulated
event we determine $L(\tb)$ for
both the electronic and nuclear recoil event classes as modeled in Sec.~\ref{sec:TimePDF}. These values are denoted $L_e(\tb)$ and $L_n(\tb)$.
We define the log-likelihood ratio, $\ln R$: 
\begin{equation}
\ln R = \ln  {L_e(\tb)}  -\ln{L_n(\tb)}.
\end{equation}
Figure~\ref{fig:MLlogR}
 shows this log-likelihood
statistic for simulated electronic and nuclear recoil events that yield 50
photoelectrons. The event is assigned to the nuclear recoil class if $\ln R$
is less than an adjustable threshold that can be varied to increase the
discrimination against electronic recoils at the cost of decreasing nuclear
recoil acceptance. 
The Monte Carlo estimate of the distribution of $\ln R$ has prominent ripples
even though the Monte Carlo estimates of
$L_n$ and $L_e$ do not. In our Monte Carlo study, we simulate
events that yield a fixed number of photoelectrons
according to PDFs with the assumption of perfect knowledge. Furthermore,
we neglect dark current noise.
In an actual experiment,
we expect that
imperfect energy resolution due to variability in 
counting statistics and dark current effects
would attenuate the ripples 
in the $\ln R$ distribution.

To illustrate the maximum likelihood method, we simulate a detector with a signal yield of 6
photoelectrons/keVee, as might be possible in a detector with full PMT coverage. We neglect dark current and we set the discrimination threshold to
accept 50\% of nuclear recoils. For comparison, we simulate prompt fraction data by assuming the number of prompt photoelectrons is a binomial
random variable with an expected value determined by the measured PDFs, and the total number of photoelectrons is assumed to be known without
error. This idealized binomial model predicts a much lower ERC than the statistical
model discussed in Sec.~\ref{sec:PromptFraction}.

\begin{figure}[ht]
  \centerline{
    \hbox{\psfig{figure=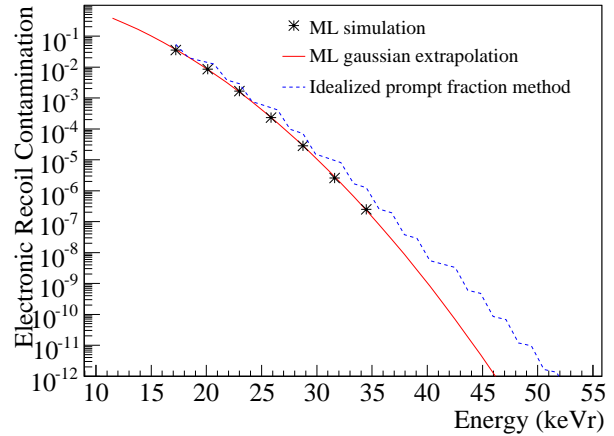,width=8.6cm,%
        clip=}}}
  \caption{(Color online) Predicted performance of maximum likelihood and
prompt ratio PSD methods for a detector yielding 6 photoelectrons/keVee (the
energy axis has been scaled from keVee to keVr by use of a constant nuclear
recoil scintillation efficiency of 0.29 as discussed in Sec.~\ref{sec:PromptFraction}.}
  \label{fig:MLPSD}
\end{figure}

The maximum likelihood
discrimination method outperforms the idealized prompt fraction method, as shown in
Fig.~\ref{fig:MLPSD}.
We also expect that the maximum likelihood method will
be more robust than the prompt fraction method to background noise.  
For the prompt fraction case, we select a threshold in prompt photoelectron space to yield
a nuclear recoil acceptance probability as close to 0.5 as possible.
Due to quantization effects, the actual nuclear
recoil acceptance probability varies about 0.5, and
there are sawtooth-like artifacts in
the prompt fraction ERC curve 
shown in 
Fig.~\ref{fig:MLPSD}.

\section{Conclusion}
Using a detector with a signal yield of 4.85 photoelectrons/keVee, we measured the scintillation time dependence of electronic and nuclear
recoils in liquid argon down to 5 keVee or 20 keVr. We developed a prompt fraction method
of PSD in liquid argon, and for a nuclear recoil acceptance level of $50\%$,
we measured a background- and statistics-limited level of ERC to
be $8.5\times10^{-6}$ between 52 and 110 keVr with no contamination events
above 69 keVr. We also developed a multibin method of PSD,
improving on the prompt fraction method by as much as an order of magnitude. With this method, we measured a background- and statistics-limited
level of ERC of $7.6\times10^{-7}$ between 52 and 110 keVr for the same nuclear
recoil acceptance of $50\%$. We modeled the observed prompt fraction data as the
ratio of two normally
distributed, correlated random variables, where we assumed $N_p$ and $N_l$ were uncorrelated; we discussed discrepancies
between observed and predicted prompt fraction results. Finally, we developed a maximum likelihood method of PSD for a detector capable
of measuring a discrete detection time for each observed photoelectron in an
event.

\begin{acknowledgments}
We thank M. Boulay, C. Jillings, B. Cai, and J. Lidgard for useful 
discussions and for developing the ratio-of-Gaussians model used in this
analysis. We acknowledge S. Seibert and J. Klein for their contributions to the
Monte Carlo simulations software package.  This work was supported by the David and Lucille Packard
Foundation, the Los Alamos Directed Research and Development Program, and the U.S. Department of Energy. 
\end{acknowledgments}
  

\end{document}